\documentclass[11pt]{article}
\usepackage{moriond,epsfig}
\usepackage{amsmath}
\usepackage{graphicx}
\usepackage{amsfonts}
\usepackage{amssymb}
\usepackage{caption}
\usepackage{natbib}

\bibpunct{[}{]}{,}{n}{,}{,}

\def\be{\begin{equation}}
\def\ee{\end{equation}}
\def\bea{\begin{eqnarray}}
\def\eea{\end{eqnarray}}

\newcommand{\s}{\scriptscriptstyle}

\begin{document}
\vspace*{3cm}
\title{NEW PHYSICS BOUNDS FROM CKM-UNITARITY \footnote{Presented at the 45th Rencontres de Moriond: Electroweak Interactions and Unified Theories, La Thuile, Italy, 6--13 Mar 2010, based on Ref. [9].}}
%ref{Cirigliano:2009wk}.}}

\author{ M. GONZ\'ALEZ-ALONSO }
\address{Departament de F\'{\i}sica Te\`orica and IFIC, Universitat de Val\`encia-CSIC,\\
Apt. Correus 22085, E-46071 Val\`encia, Spain \\
Department of Physics, University of Wisconsin-Madison, Madison, WI 53706 USA.}

\maketitle\abstracts{
Using an effective field theory approach, we study the new physics (NP) corrections to muon and beta decays and their effects on the extractions of $V_{ud}$ and $V_{us}$. Assuming nearly flavor blind NP interactions we find that the CKM-unitarity test is the only way to expose NP. The four short-distance operators that can generate a deviation are strongly constrained by the phenomenological bound $|V_{ud}|^2 + |V_{us}|^2 + |V _{ub}|^2 - 1 = (-1 \pm 6) \times 10^{-4}$, corresponding to an effective scale $\Lambda > 11$ TeV (90\% CL). Depending on the operator, this constraint is at the same level or better than that generated by the Z pole observables.
}

\section{Introduction}\label{sec:Introduction}

In the last years there have been a continuous advance in the theoretical description of semileptonic kaon decays, both in the lattice sector~\cite{lattice} and using analytical approaches~\cite{analytical}. These improvements, in combination with new experimental measurements~\cite{Amsler:2008zzb}, make these decays a deep probe of the nature of weak interactions~\cite{Antonelli:2008jg,Antonelli:2010yf}. In particular, the elements $V_{ud}$ and $V_{us}$ of the  Cabibbo-Kobayashi-Maskawa (CKM)~\cite{CKM} quark mixing matrix are known with an accuracy below the percent level: $V_{ud} = 0.97425(22)$~\cite{Hardy:2008gy} and $V_{us} =0.2253(9)$~\cite{Antonelli:2010yf}. These precise determinations can be used to test the CKM unitarity condition \footnote{$V_{ub} \sim 10^{-3}$ contributes negligibly to this relation.} 
\bea
\Delta_{\rm CKM} \equiv  |V_{ud}^{(\rm pheno)}|^2+|V_{us}^{(\rm pheno)}|^2+|V_{ub}^{(\rm pheno)}|^2 \ - \ 1 = (-1 \pm 6)\times 10^{-4},
\eea
or equivalently, the quark-lepton universality. Assuming that new physics contributions scale as $\alpha/\pi   (M_W^2/\Lambda^2)$, the unitarity test probes energy scales  $\Lambda$ on the order of the TeV, which will be directly probed at the LHC.

While the consequences of these unitarity tests on Standard Model  (SM) extensions have been considered in some explicit scenarios~\cite{modeldependent}, a model-independent analysis was missing until recently~\cite{Cirigliano:2009wk}. In order to perform such an analysis, the main idea is to study in a model-independent effective theory setup new physics  contributions to low energy charged-current (CC) processes, in such a way that we can assess  in a fairly general way the impact of semileptonic processes in constraining and discriminating  SM extensions. We shall pay  special attention to purely leptonic and semileptonic decays of light hadrons used to extract the CKM elements $V_{ud}$ and $V_{us}$.

\section{Weak scale effective lagrangian}
\label{sect:weakscale}

In order to analyze in a model-independent framework NP contributions to both beta decays and electroweak precision observables (EWPO) we take the SM (including the Higgs) as the low-energy limit of a more fundamental theory, and more specifically we assume that: (i) there is a gap between the weak scale $v$ and the NP scale $\Lambda$ where new degrees of freedom appear; (ii) the NP at the weak scale is weakly coupled, so the electroweak (EW) gauge symmetry is linearly realized; (iii) the violation of total lepton and baryon number is suppressed by a scale much higher than $\Lambda$. These assumptions lead us to an effective non-renormalizable lagrangian~\cite{LEFF}:
\bea
\label{eq:EFT}
{\cal L}^{(\rm{eff})} &=& {\cal L}_{\rm{SM}} + \frac{1}{\Lambda} {\cal L}_5 + \frac{1}{\Lambda^2} {\cal L}_6 + \frac{1}{\Lambda^3} {\cal L}_7 + \ldots
\eea
where ${\cal L}_n = \sum_i \alpha^{(n)}_i O_i^{(n)}$, being ${\cal O}_i^{(n)}$ local gauge-invariant operators of dimension $n$ built out of SM fields. It can be shown that under the above assumptions, there are no corrections at dimension five, whereas seventy-seven operators appear at dimension six~\cite{Cirigliano:2009wk,LEFF}, where we truncate the expansion. For the sake of consistency we will work at linear order in the NP corrections.

For the EWPO {\it and} beta decays it can be shown that we only need a twenty-five operator basis, with twenty-one $U(3)^5$ invariant and four non-invariant\footnote{We refer to the $U(3)^5$ flavor symmetry of the SM gauge lagrangian (the freedom to make $U(3)$ rotations in family space for each of the five fermionic gauge multiplets).} (we will see the usefulness of this separation later). Nine of those operators contribute to the beta and muon decays, being the following five the only $U(3)^5$-invariant:
\begin{eqnarray}
 && O_{ll}^{(1)}=\frac{1}{2} (\overline{l} \gamma^\mu l) (\overline{l} \gamma_\mu l)
 ~~~~~~~~~~~~~~~~~~O_{ll}^{(3)} = \frac{1}{2} (\overline{l} \gamma^\mu \sigma^a l) (\overline{l} \gamma_\mu \sigma^a l),  \label{eq:oll} \\
 && O_{l q}^{(3)}= (\overline{l} \gamma^\mu \sigma^a l) (\overline{q} \gamma_\mu \sigma^a q), \label{eq:olq} \\
 && O_{\varphi l}^{(3)}=\! i (h^\dagger\!D^\mu \sigma^a \!\varphi)(\overline{l} \gamma_\mu \sigma^a l)+\!{\rm h.c.}, \label{eq:ohl}
 ~ O_{\varphi q}^{(3)} =\! i (\varphi^\dagger\!D^\mu \sigma^a \!\varphi)(\overline{q} \gamma_\mu \sigma^a\! q)+\!{\rm h.c.}, \label{eq:ohq}~.
 %\\
%&& O_{qde} = (\overline{\ell} e) (\overline{d} q)+ {\rm h.c.}, \label{eq:oqde} \\
%&& O_{l q} = (\bar{l}_a e)\epsilon^{ab}(\bar{q}_b u)+ {\rm h.c.} \label{eq:olq2}
%~~~~~~~~~~~ O^t_{l q} = (\bar{l}_a\sigma^{\mu\nu}e)\epsilon^{ab}(\bar{q}_b\sigma_{\mu\nu}u)+ {\rm h.c.} \\
%&& O_{\varphi \varphi} = i(\varphi^T \epsilon D_\mu \varphi) (\overline{u}\gamma^\mu d)+ {\rm h.c.}~. \label{eq:ohh}
\end{eqnarray}

\section{Effective lagrangian for $\mu$ and quark $\beta$ decays}
\label{sect:gevscale}
Deriving the low-energy effective lagrangian that describes the muon and beta decays we find \cite{Cirigliano:2009wk}
\bea
%{\cal L}_{\mu \to e \bar{\nu}_e \nu_\mu}
{\cal L}_{\mu}
= \frac{-g^2}{2 m_W^2}  \! \! \! \!\! \! \! \! \! \! \! \!&& \Bigg[ \left(1 +
\tilde{v}_L
\right) \cdot \bar{e}_{\s{L}} \gamma_\mu \nu_{e\s{L}}
\ \bar{\nu}_{\mu \s{L}} \gamma^\mu \mu_{\s{L}}  \ + \
\tilde{s}_R
\cdot \bar{e}_{\s{R}} \nu_{e\s{L}} \ \bar{\nu}_{\mu \s{L}} \mu_{\s{R}} \Bigg] ~+~ h.c.~, ~
\label{eq:leffmu}\\
%
%\label{eq:A}
%\tilde{v}_L &= & 2~[\hat{\alpha}_{\varphi l}^{(3)}]_{11+22^*} - [\hat{\alpha}_{ll}^{(1)}]_{1221} - 2 [\hat{\alpha}_{ll}^{(3)}]_{1122-\frac{1}{2}(1221)}\\
%
%\label{eq:B}
%\tilde{s}_R &=& +2 [\hat{\alpha}_{le}]_{2112}~,\\
%{\cal L}_{d_j \to u_i \ell^- \bar{\nu}_\ell}
{\cal L}_{d_j}
= \frac{-g^2}{2 m_W^2} \, V_{ij}  \! \! \! \!\! \! \! \! \! \! \! \!&&\Bigg[
 \Big(1 + [v_L]_{\ell \ell ij} \Big) \ \bar{\ell}_L \gamma_\mu \nu_{\ell L} \ \bar{u}_L^i \gamma^\mu d_L^j
 \ + \ [v_R]_{\ell \ell ij} \ \bar{\ell}_L \gamma_\mu \nu_{\ell L} \ \bar{u}_R^i \gamma^\mu d_R^j
\nonumber\\
&&+ [s_L]_{\ell \ell ij} \ \bar{\ell}_R \nu_{\ell L} \ \bar{u}_R^i d_L^j
\ + \ [s_R]_{\ell \ell ij} \ \bar{\ell}_R \nu_{\ell L} \ \bar{u}_L^i d_R^j
\nonumber \\
&&+ [t_L]_{\ell \ell ij} \ \bar{\ell}_R \sigma_{\mu \nu} \nu_{\ell L} \ \bar{u}_R^i \sigma^{\mu \nu} d_L^j
\Bigg]~+~h.c.~.
\label{eq:leffq} 
%\\
%\label{eq:beta01}
%V_{ij}  \cdot \left[v_{L}\right]_{\ell \ell i j} &=& 2 \, V_{ij} \, \left[\hat{\alpha}_{\varphi l}^{(3)}\right]_{\ell\ell} + 2 \, V_{im} \left[\hat{\alpha}_{\varphi q}^{(3)}\right]_{jm}^*- 2\, V_{im} \left[\hat{\alpha}_{l q}^{(3)}\right]_{\ell\ell mj} \\
%V_{ij} \cdot \left[v_R\right]_{\ell \ell ij } &=& - \left[\hat{\alpha}_{\varphi \varphi}\right]_{ij} \\
%V_{ij} \cdot \left[s_L\right]_{\ell \ell ij } &=& - \left[\hat{\alpha}_{l q}\right]_{\ell\ell ji}^* \\
%V_{ij} \cdot \left[s_R\right]_{\ell \ell ij} &=& - V_{im}\left[\hat{\alpha}_{qde}\right]_{\ell\ell jm}^* \\
%V_{ij} \cdot \left[t_L\right]_{\ell \ell ij } &=& - \left[\hat{\alpha}^t_{l q} \right]_{\ell\ell ji}^* ~.
%\label{eq:beta1}
\eea
The effective couplings $\tilde{v}_L,\tilde{s}_R,v_{L,R},s_{L,R}$, and $t_L$ encode information on interactions beyond the SM  \cite{Cirigliano:2009wk} and are of order $v^2/\Lambda^2$, where $v$ is the SM Higgs expectation value. 
%For example we have
%\bea
%\tilde{v}_L &= & 2~[\hat{\alpha}_{\varphi l}^{(3)}]_{11+22^*} - [\hat{\alpha}_{ll}^{(1)}]_{1221} - 2 [\hat{\alpha}_{ll}^{(3)}]_{1122-\frac{1}{2}(1221)}
%\eea
%The coupling vL receives contributions from three gauge invariant weak scale operators (gauge boson-quark vertex correction, gauge boson-lepton vertex correction and contact four fermion) while the other couplings are in one-to-one correspondence with gauge invariant four-fermion operators at the weak scale.
%

\section{Flavor structure of the effective couplings}
\label{sect:flavor}
%
%Using the general effective lagrangians (\ref{eq:leffmu}) and (\ref{eq:leffq}) for CC transitions, one can calculate the deviations from SM predictions in various SL decays and in principle a rich phenomenology is possible, because $V_{ud}$ and $V_{us}$ can be determined with high precision through different channels.
%
So far we have not made any assumption about the flavor structure of the new physics, but given that flavor changing neutral current (FCNC) processes forbid generic structures if $\Lambda \sim {\rm TeV}$, it is convenient to organize the discussion in terms of perturbations around the $U(3)^5$ flavor symmetry limit, where no problem arises with FCNC. In this $U(3)^5$-limit the expressions greatly simplify: the effective couplings $\tilde{s}_R,v_{R},s_{L,R}$, and $t_L$ vanish and all the NP effects can be encoded into the following redefinitions
\bea
G_F^\mu &=& (G_F)^{(0)} \, \left(1+\tilde{v}_L  \right)= (G_F)^{(0)} \, \left(1 + 4 \, \hat{\alpha}_{\varphi l}^{(3)} - 2 \,  \hat{\alpha}_{ll}^{(3)} \right)~, \\
G_F^{\rm SL}&=& (G_F)^{(0)} \, \left(1+v_L \right)= (G_F)^{(0)} \, \left( 1 + 2 \left( \hat{\alpha}_{\varphi l}^{(3)} + \hat{\alpha}_{\varphi q}^{(3)}- \hat{\alpha}_{l q}^{(3)}\right)\right)~,
\eea
where $G_F^{(0)} = g^2/(4 \sqrt{2} m_W^2)$. Consequently we will have
\bea
\label{eq:vphenoFB}
V_{ij}^{(\rm pheno)}
&=& \frac{V_{ij}\,G_F^{\rm SL}}{G_F^\mu} = V_{ij} \left[1 +2\,\left( \hat{\alpha}_{ll}^{(3)} -\hat{\alpha}_{lq}^{(3)} -\hat{\alpha}_{\varphi l}^{(3)} +\hat{\alpha}_{\varphi q}^{(3)}\right)\right]~,
\eea
as phenomenological values of $V_{ud,us}$, independently of the channel used to extract them. Therefore the only way to expose NP contributions is to construct universality tests ($\Delta_{\rm CKM}\neq 0$), in which the absolute normalization of $V_{ij}$ matters. In our framework we have
\be
\Delta_{\rm{CKM}}
=       4 \, \left( \hat{\alpha}_{ll}^{(3)} -\hat{\alpha}_{l q}^{(3)} - \hat{\alpha}_{\varphi l}^{(3)} + \hat{\alpha}_{\varphi q}^{(3)}  \right)~.
\label{eq:dckmnp}
\ee
In specific SM extensions, the $\hat{\alpha}_i$ are functions of the underlying parameters. Therefore, through the above relation one can work out the constraints of quark-lepton universality tests on any weakly coupled SM extension.

The Minimal Flavor Violation hypothesis requires that $U(3)^5$ symmetry is broken in the underlying model only by structures proportional to the SM Yukawa couplings~\cite{MFV}, and structures generating neutrino masses~\cite{Cirigliano:2005ck}, and therefore the coefficients parameterizing deviations from the $U(3)^5$-limit are highly suppressed \cite{Cirigliano:2009wk}. Consequently we expect the conclusions of the previous subsection to hold, with the elements $V_{ij}$ receiving a common dominant shift plus suppressed channel-dependent corrections.

In a more general framework the situation can be different because the channel-dependent shifts to $V_{ij}$ could be appreciable and $\Delta_{\rm CKM}$ would depend on the channels used. Work in this direction is in progress.

\section{$\Delta_{\rm CKM}$ versus precision EW measurements}
\label{sec:InvaraintAnalysis}

The four operators that contribute to $\Delta_{\rm CKM}$ in the limit of approximate $U(3)^5$ invariance also contribute to the different EWPO~\cite{Han:2004az}, together with the remaining seventeen operators that make up the $U(3)^5$ invariant sector of our TeV scale effective lagrangian. Han and Skiba~\cite{Han:2004az} studied the constraints on the same set of twenty-one $U(3)^5$ invariant operators from the EWPO, performing a global fit, and from this work we have the following indirect bound on $\Delta_{\rm CKM}$
\be
 - 9.5 \times 10^{-3}  \ \leq \ \Delta_{\rm CKM} \ \leq \ 0.1 \times 10^{-3} \qquad (90\% \ {\rm C.L.})~.
\ee
Comparing with the direct experimental limit, $|\Delta_{\rm CKM}| \leq 1. \times 10^{-3}$ ($90\%$ C.L.), we see that EWPO leave room for a sizable violation of unitarity and consequently we have to include the direct $\Delta_{\rm CKM}$ constraint in the global fit to improve the bounds on NP-couplings. It has been shown \cite{Cirigliano:2009wk} that the main effect of this addition is to strengthen the constraints on $O_{l q}^{(3)}$.

\begin{figure}[t]
%\hspace{0.3cm}
\centering \includegraphics[width=0.85\textwidth]{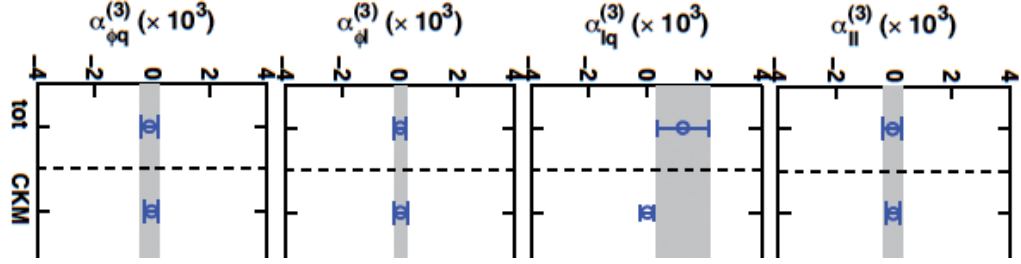}
\captionsetup{justification=justified,font=small}
\caption{$90 \, \%$ C.L. regions in the single operator analysis. The first row displays the constraint from EWPO and the second row those coming only from $\Delta_{\rm CKM}$.}
\label{fig:ZoomedIndOpConst}
\end{figure}

In Fig.~\ref{fig:ZoomedIndOpConst} we show the bounds if we assume a single operator dominance. For all the CKM-operators the direct $\Delta_{\rm CKM}$ measurement provides competitive constraints and in the case of $O_{lq}^{(3)}$ the improvement is remarkable.

\section{Conclusions}
\label{sec:Conclusions}

In an effective field theory framework and assuming nearly $U(3)^5$-invariant NP interactions, it has been shown that the extraction of $V_{ud,us}$ is channel independent and the only NP probe is $\Delta_{\rm CKM}$, that receives contributions from four short distance operators: $O_{ll,lq,\varphi l, \varphi q}^{(3)}$.

We have shown that the CKM-unitarity (first row) test provide constraints on NP that currently cannot be obtained from other EW precision tests and collider measurements. The $\Delta_{\rm CKM}$ constraint bounds the effective NP scale of all four CKM-operators to be $\Lambda > 11$ TeV (90 \% C.L.), what for $O_{l q}^{(3)}$ is five times stronger than EWPO-bound. Equivalently, if $V_{ud}$ and $V_{us}$ move from their current central values, EWPO data would leave room for sizable deviations from CKM-unitarity.

\section*{Acknowledgments}
Work supported by the EU RTN network FLAVIAnet [MRTN-CT-2006-035482] and MICINN, Spain [FPU No. AP20050910, FPA2007-60323 and CSD2007-00042 -CPAN-].

\end{document}